# Setup and performance of focal-plane characterization benches for high precision astrometry

Fabrice Pancher*[a], Sébastien Soler[a], Fabien Malbet[a], Hugo Rousset[a], Manon Lizzana[a,b,c], Alain Leger[d], Thierry Lepine[e], Florence Ardellier-Desages[f], Jérôme Amiaux[f], Pierre-Olivier Lagage[f]

[a]Université Grenoble Alpes, CNRS, IPAG, Grenoble, France; [b]Centre National d'Études Spatiales, Toulouse, France; [c]Pyxalis, Moirans, France ; [d]Université Paris-Saclay, CNRS, CNES, IAS, Gif-sur-Yvette, France ; [e]Université Jean Monnet - Saint-Etienne, IOGS, CNRS, Laboratoire Hubert Curien, Saint-Etienne, France ; [f]Univ. Paris-Saclay, Univ. Paris Cite, CEA, CNRS, AIM

## ABSTRACT

Detecting Earth-like exoplanets requires micro-arcsecond astrometry, which relies on a precisely calibrated focal plane equipped with high-resolution detectors. Accurate calibration depends on characterizing detector imperfections and systematic effects. Optical benches have been developed to assess the sensor's behavior: an integrating-sphere setup for electro-optical parameters, an interferometric bench for pixel-level geometry, and a star-field projection system for intrinsic distortion. This work presents the characterization benches built for a 46-Megapixels (MP) imaging sensor, based on the same technology as the four 220-MP detectors foreseen for the final focal-plane, and the performance metrics obtained from these setups in the context of NASA's HWO and ESA's M-class missions. Upgrades will bring the test environment closer to operational conditions: a second 46-MP sensor will be added for multi-detector tests, plus an adjustable-angle fringe-projection system and improved alignment.

## 1. INTRODUCTION

Differential astrometry enables high-precision measurements of the position and motion of celestial objects by comparing a target to reference stars within the same field of view. This technique allows the detection of Earth-like exoplanets around nearby stars [1], where the gravitational influence of an orbiting planet induces a small stellar motion at the level of $\sim 10^{-5}$ pixel [1-4]. While missions such as Gaia have demonstrated astrometry at the micro-arcsecond level, future concepts like Theia, targeting ~0.3 µas corresponding to $\sim 5\times10^{-6}$ pixel, and astrometric modes envisioned for the Habitable Worlds Observatory [5] (HWO) mission targeting the $\sim 5\times10^{-5}$ pixel level, are situated in the sub-micro-arcsecond performance domain. Achieving this level of precision requires controlling instrumental error sources well below the pixel scale, including optical distortion, detector geometry, and inter/intra-pixel response variations, down to the $10^{-5}$ pixel level. These requirements motivate the development of large-format CMOS focal planes, which may be composed of multiple high-resolution detectors tiled to form a single focal plane.

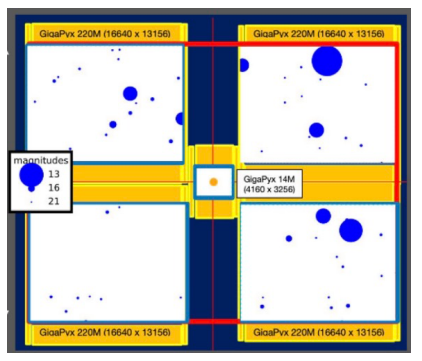

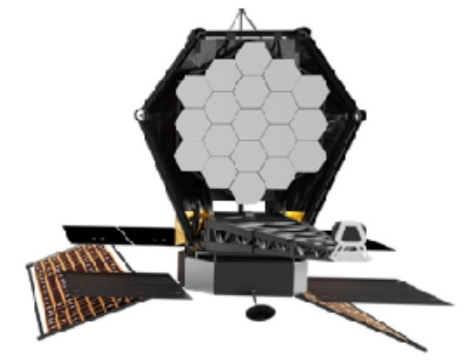

Figure 1. Predicted focal plane assembly for the HWO satellite (4x 220-MP +1x 13-MP central sensors)

*fabrice.pancher@univ-grenoble-alpes.fr; phone +33 4 76 14 37 04; https://ipag.osug.fr

In this context, detector geometry characterization and focal plane calibration are limiting factors, requiring dedicated calibration approaches [6] beyond classical methods. This paper presents the calibration benches developed and currently used for the characterization of the 46-MP CMOS detectors from the Pyxalis Gigapyx [7] family.

These detectors were selected following an evaluation of several available solutions, as they best satisfy the requirements of the HWO mission, including the resolution planned for future Gigapyx sensors, the absence of microlenses, pixel size, and frame rate. In addition, the Gigapyx 46-MP sensor showed good performances during radiation testing, making it a good candidate for space missions.

| Company | Model | Type | Resolution | Megapixels | Pixel size(µm) | Frame/s | radiation test | micro-lens |
|---|---|---|---|---|---|---|---|---|
| Canon (Japan) | 120MXS | CMOS | 13272 x 9176 | 122 | 2,20 | 9,40 | NO | YES |
| Canon (Japan) | LI8020SA | CMOS | 19568 x 12588 | 250 | 1,50 | 5 | NO | YES |
| Canon (Japan) | LI8030SA | CMOS | 24592 x 16704 | 410 | 1,50 | 8 | NO | YES |
| Teledyne (USA) | CCD290-99 | CCD | 9216 x 9232 | 85 | 10,00 | 3M | YES | NO |
| Teledyne (USA) | CCD231-C6 | CCD | 6144 x 6160 | 38 | 15,00 | 3M | YES | NO |
| Sony (Japan) | IMX411ALR | CMOS | 14208 x 10656 | 151 | 3,76 | 6 | NO | YES |
| Sony (Japan) | IMX811-AAMR | CMOS | 19240 x 12840 | 247 | 2,81 | 12 | NO | YES |
| Sony (Japan) | IMX661-AAMR | CMOS | 13472 x 9568 | 129 | 3,45 | 21 | NO | YES |
| Gpixel (China) | GSENSE 1081 BSI | CMOS | 8900 x 9120 | 81 | 10,00 | 0,34 |  | NO |
| Pyxalis (France) | Gigapyx 220 | CMOS | 16640 x 13156 | 220 | 4,40 | 13 | YES | NO |

Table 1. Overview of different image sensors.

The paper also describes an on-going upgrade of the bench designed to accommodate either two 46-MP sensors or a single 220-MP detector, providing a test configuration that is more representative of the future HWO focal plane architecture.

## 2. EXPERIMENTAL BENCH FOR DETECTOR CHARACTERIZATION UNDER FLAT-FIELD ILLUMINATION

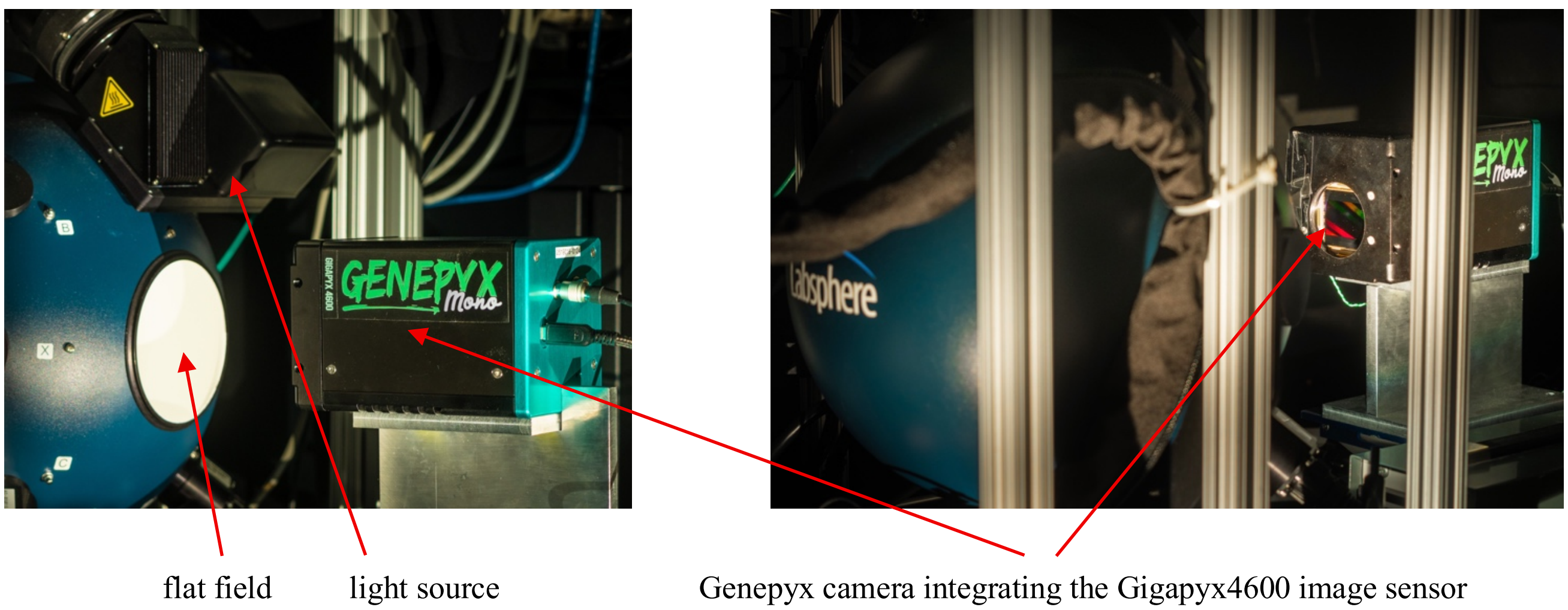

Figure 2. Flat field calibration bench.

In the initial phase of the calibration activities, a single 46-MP Gigapyx CMOS detector was used. The sensor is integrated into readout electronics developed by Pyxalis and, while not fully optimized in terms of performance, it provides a sufficient platform for the present characterization activities. A Python library was developed on top of the Pyxalis Software

Development Kit (SDK) to provide a unified API, enabling the detector and the other controllable bench components to be operated from a common Python interface. This facilitates synchronized operation and automated data acquisition. The first step of the characterization consists of measuring the standard detector performance parameters using a calibrated flat-field illumination. For this purpose, an integrating sphere (Labsphere HeliosPlus family, USLR-D12F-NDNN-P) provides a uniform illumination over the detector surface, with a specified spatial luminance uniformity better than ±1% across the exit port and a maximum output luminance of 50,000 cd/m². The sphere is equipped with three independently switchable halogen lamps and a regulated motorized adjustable diaphragm, allowing fine control of the output flux.

Although operated via its proprietary software, a dedicated control library has been developed to enable remote operation and full integration of the integrating sphere control within the Python-based unified API. This allows synchronized control of illumination conditions and detector acquisition, ensuring reproducible measurements for standard detector characterization tasks, including linearity, gain estimation, dark current, and noise performance.

## 3. TEST BENCH FOR PIXEL AND OPTICAL DISTORSION CALIBRATION

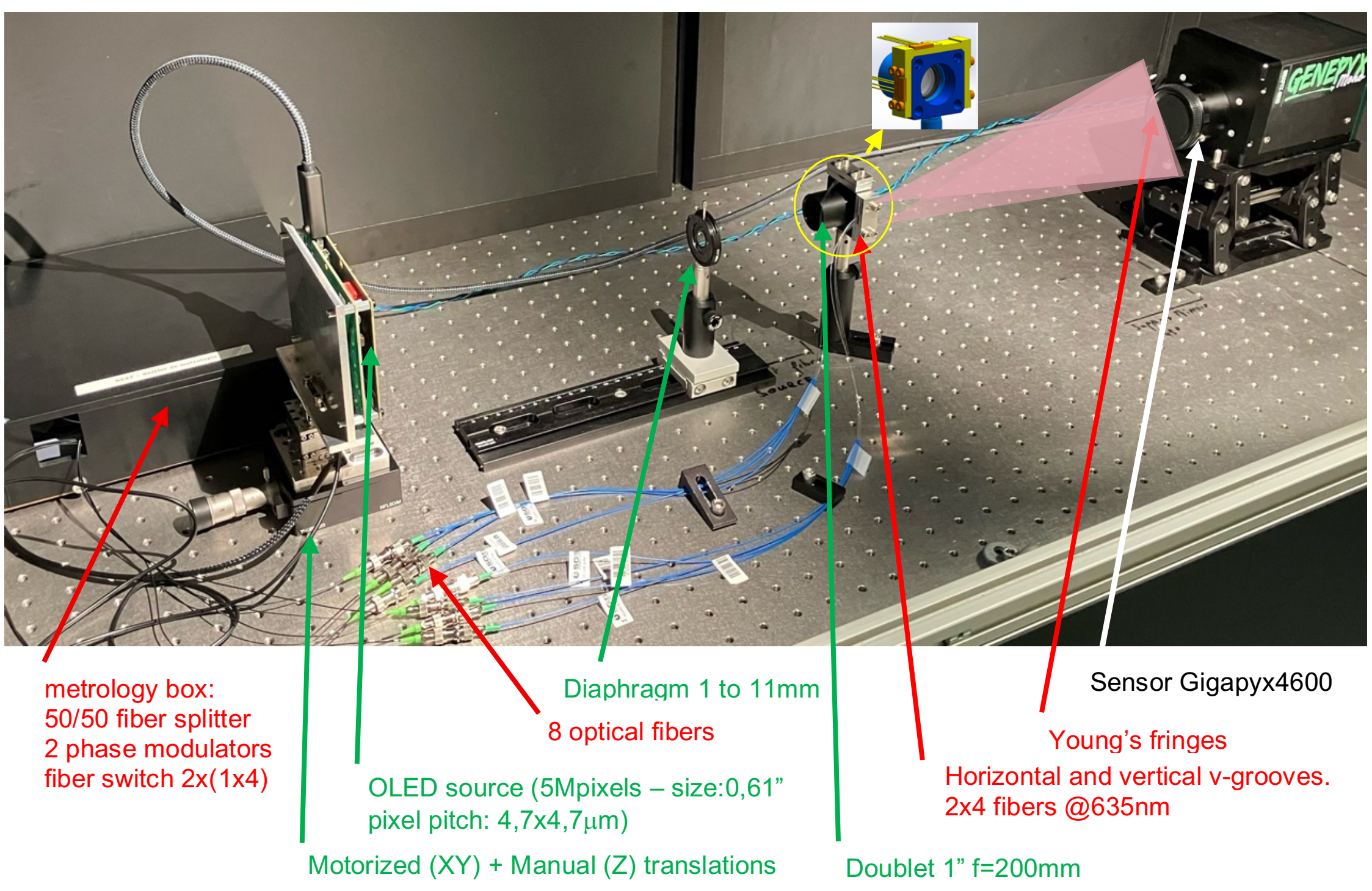

Figure 3. Interferometric and distortion calibration bench.

The same bench is used for both pixel geometry and distortion characterization. As illustrated in Figure 3, the pixel geometry measurements are performed through a pair of fibers generating Young’s fringes, mounted on the central optical mount of the distortion bench.

## Experimental Setup for Distortion

Reference stars with known positions from the Gaia mission catalogue can be used to derive a mathematical model [8] describing the optical distortion across the instrument. This model establishes the relationship between the measured detector coordinates and the true object positions, enabling the correction of errors introduced by the optical system. To experimentally validate this calibration approach, a dedicated laboratory distortion bench has been developed.

The bench reproduces a stellar field by displaying a programmable star pattern with known coordinates using a 5-MP OLED screen as the source. The pattern is projected onto the detector through a 2f–2f optical configuration employing an achromatic doublet located at the midpoint between the source and the sensor, designed such that each source pixel is imaged onto at least one pixel of the detector. Controlled distortion is introduced by manually translating a diaphragm along the optical axis and manually adjusting its aperture. This allows generating deviations between the known source coordinates and the corresponding coordinates on the detector. The resulting position offsets are used to fit a polynomial distortion model over the detector. This function can be applied to subsequent measurements to reconstruct the undistorted focal-plane coordinates and recover the true positions of observed sources.

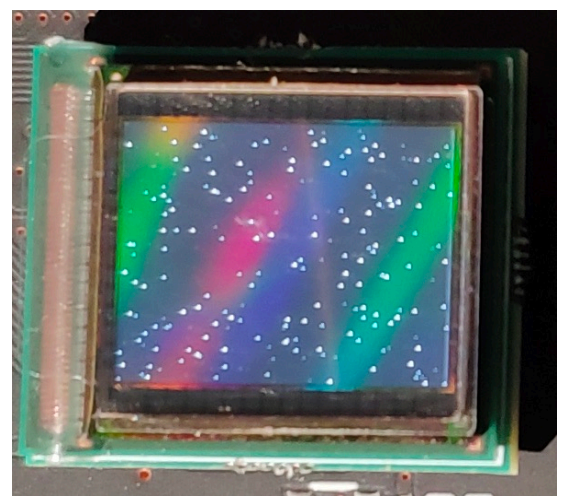

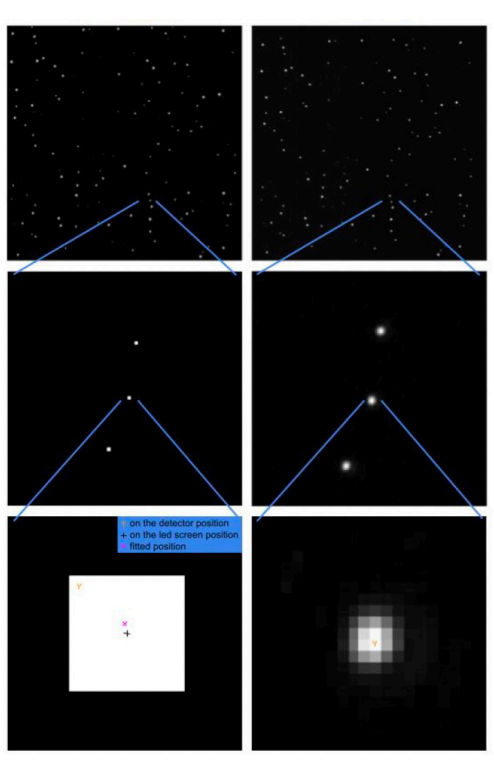

Figure 4. Left: star field shown on the 5-MP OLED screen.
Right-column 1: zoom of the star field on shown on the screen.
Right-column 2: zoom of the star field captured on the sensor.

## Experimental setup for pixel geometry calibration [9]

In addition to optical distortion calibration, high-precision differential astrometry requires an accurate characterization of the detector pixel geometry. At the sub-micro-arcsecond level, the assumption of a perfectly regular pixel grid is no longer valid, as individual pixel centers can exhibit small position offsets resulting from manufacturing tolerances. These offsets must be calibrated with an accuracy significantly better than one thousandth of a pixel. To address this requirement, an interferometric bench has been developed to measure the positions of pixel centers on the detector.

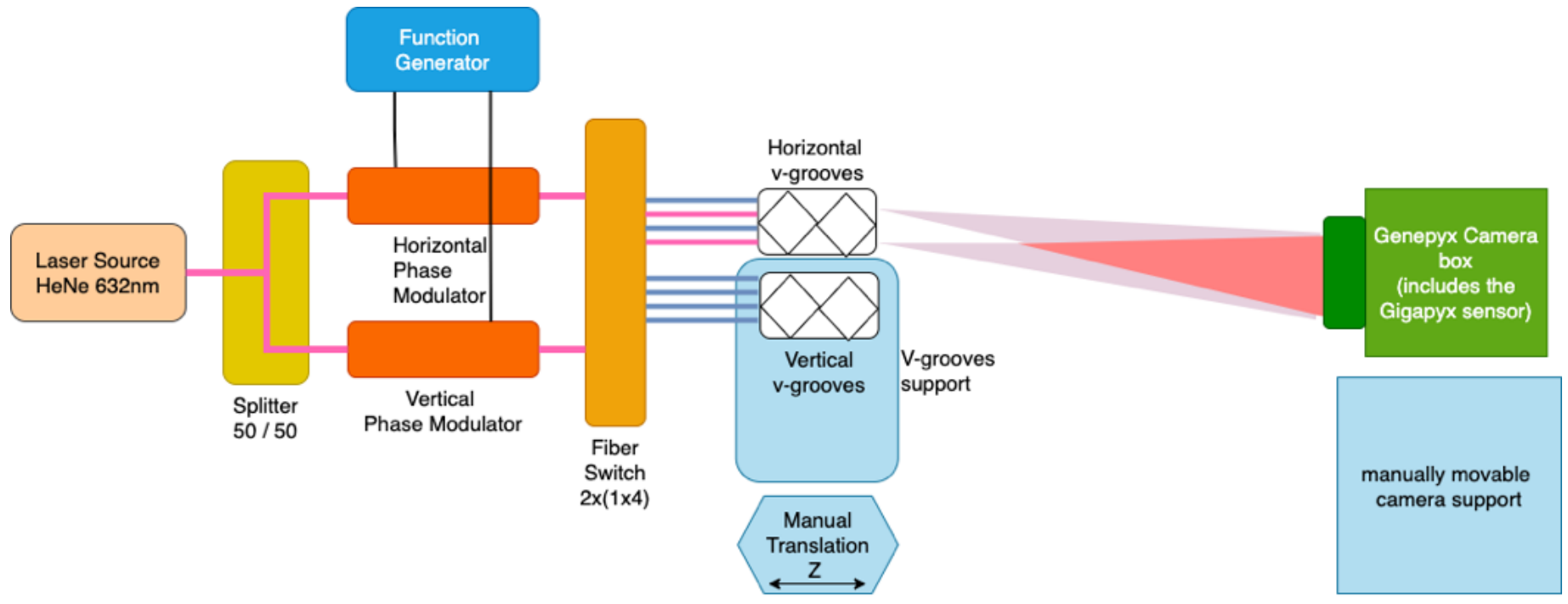

Figure 5. Interferometric calibration bench concept

The calibration principle relies on the projection of Young's fringes [10] generated by the interference of two coherent optical beams. An HeNe laser operating at 632.8 nm is split into two optical paths, each passing through a lithium-niobate phase modulator. The applied phase modulation produces a controlled displacement of the interference fringes across the detector, allowing the temporal response of each pixel to be measured. The two optical beams are routed through a 2-inputs / 8-outputs fibers switch toward two V-grooves, one horizontal and one vertical, each including four single-mode fibers. The switch allows to selects a pair of fibers within either the horizontal or the vertical V-groove, enabling different fiber separations and consequently different interference fringe spatial frequencies. The position of each pixel center is estimated by comparing the measured fringe phase at each pixel with the expected phase distribution of the interference pattern, enabling sub-pixel accuracy. This calibration provides a precise pixel geometry map that can be used to correct pixels centroiding errors.

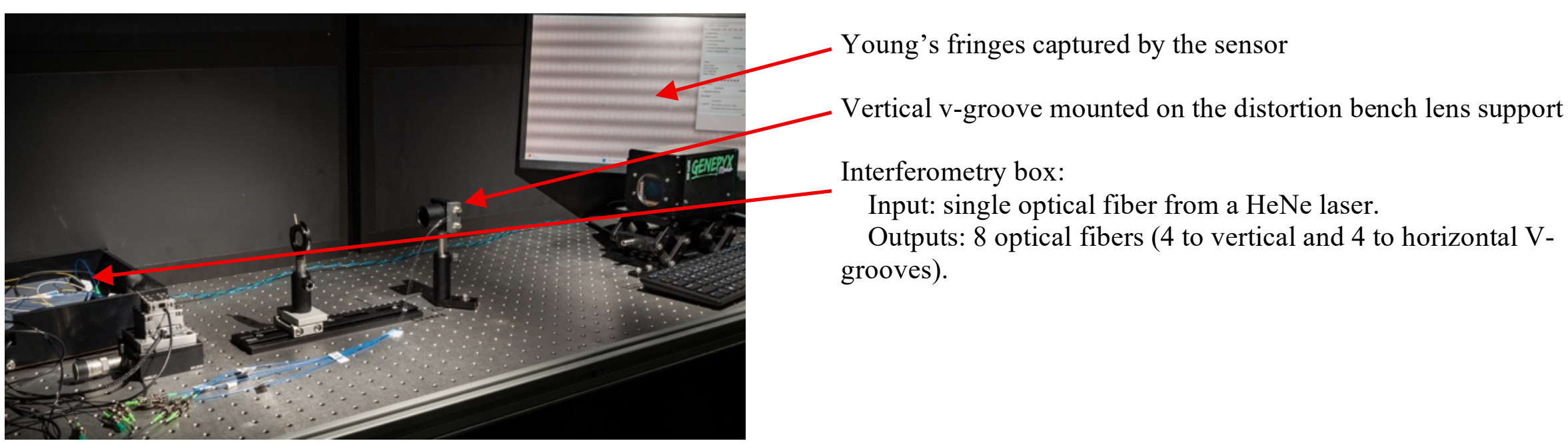

Figure 6. Interferometric bench with fringes captured on the sensor

# 4. CONTROL ARCHITECTURE

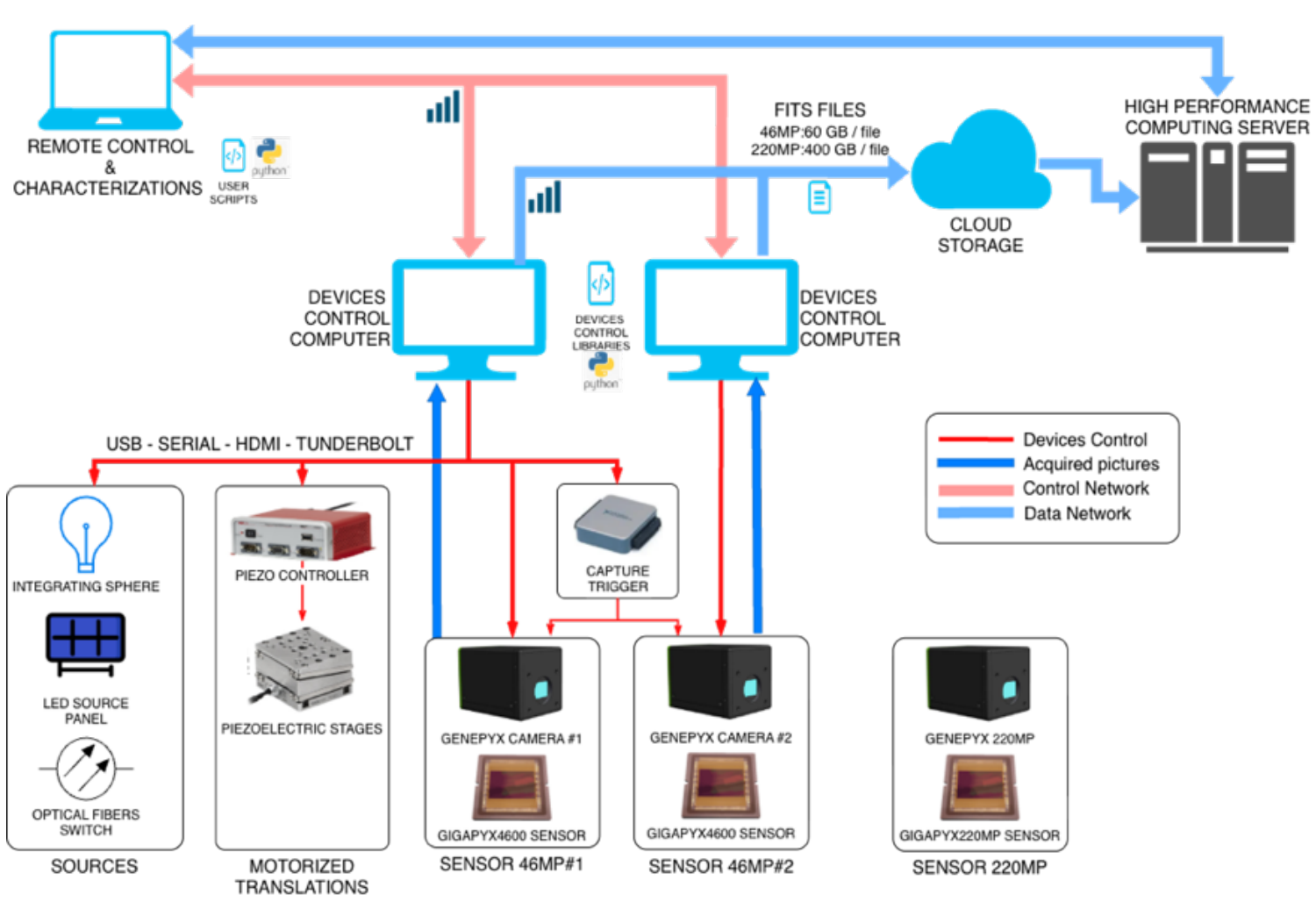

Figure 7. Control architecture overview

The control architecture is organized around two dedicated computers. The first operates the instrumentation used for detector characterization. Depending on the characterization being performed, it controls several devices, including an integrating sphere for flat-field measurements, an LED panel used to generate star fields for distortion measurements, piezoelectric translation stages for motorized source positioning, and an optical switch that select the fiber pairs used to generate Young's interference fringes. Phase modulators are currently driven by an external voltage generator and are operated manually. Future upgrades of the interferometric bench will introduce additional motorized degrees of freedom, including rotation of the V-groove assembly and tip, tilt, and rotational alignment of one GIGAPYX-46MP detector relative to the other.

These campaigns typically require several thousand image acquisitions and can last for tens of hours. A set of Python libraries has been developed to automate the characterization procedures by providing a unified interface to all hardware components. These libraries allow users to control different hardware components using the same Python scripts while abstracting device-specific communication protocols. Due to current limitations of the camera driver, which does not support simultaneous operation of multiple detectors on a single computer, the dual-camera GIGAPYX-46MP configuration requires a second computer dedicated to the control of the second camera. Image acquisition is synchronized using an external trigger system controlled by the first computer, ensuring simultaneous exposures on both sensors.

All acquired data are stored on a shared storage server accessible from both acquisition computers and from a dedicated high-performance computing workstation. This infrastructure enables efficient processing of the large datasets generated during detector characterization campaigns. Such capabilities are required given both the number of acquired frames and their size, with a single image representing approximately 100 MB for the 46-MP detector and up to 450 MB for the future 220-MP sensor.

## 5. FIRST RESULTS

| Property | Pyxalis | IPAG |
|---|---|---|
| read out noise | 16 $e^-$ | 12 $e^-$ |
| linearity error | 2% | 0.90% |
| bad pixels | $\leq 0.1\%$ | 0.02 % |
| response non uniformity | $\leq 2\%$ | 1.3% |
| dark current (at 35°C) | 24 $e^-$/s | 36 $e^-$/s |
| saturation charge | >50 k$e^-$ | 56 k$e^-$ |
| percentage of defective pixels | - | 0.0004% |
| inter-pixel capacitance | - | $\leq 3\%$ |

Table 2: Gigapyx 4600 characterization results

The experimental characterization of the GIGAPYX4600 is consistent with Pyxalis specifications. The sensor shows improved linearity (0.90 ± 0.18% error vs 2% specified), reduced readout noise (12 ± 3 $e^-$ vs. 16 $e^-$), and controlled PRNU (≤ 1.3% vs. ≤ 2%). The measured dark current (36 ± 20 $e^-$/s at 35°C) slightly exceeds the specified value (24 $e^-$/s). The device demonstrates suitability for high-precision astrometry. Inter-pixel capacitance (IPC < 3%) remains a limiting factor but is mitigated through sensor configuration and post-processing.

Preliminary pixel maps are consistent with the detector electrical layout provided by the manufacturer. Measured pixel position deviations are within approximately −0.5% to +0.5% of a pixel (~44 nm), although repeatability and accuracy are still under investigation.

## 6. DESIGN OF AN UPGRADED INTERFEROMETRIC CALIBRATION BENCH

A major upgrade of the interferometric calibration bench is currently being designed to better reproduce HWO focal-plane. The upgraded facility will remain modular and will support either a pair of adjacent 46-MP Gigapyx detectors or a single 220-MP detector. Although the bench will not replicate the complete focal-plane assembly, the dual-detector configuration provides a representative platform for studying the multi-detector architecture planned for HWO, where four 220-MP detectors will surround a central 14-MP detector. This platform will enable the validation of detector co-alignment, relative calibration strategies, and metrology concepts applicable to the future focal-plane implementation.

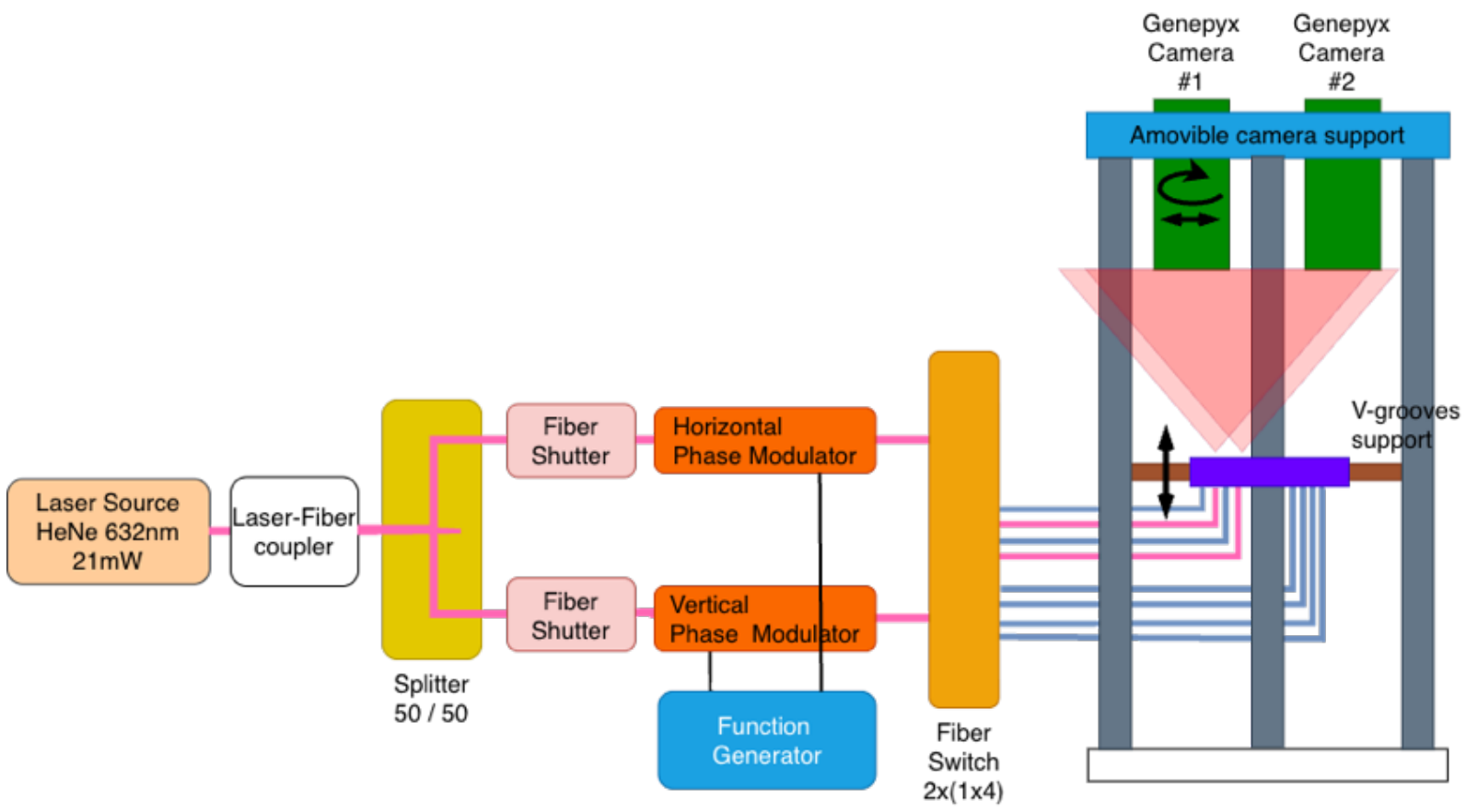

Figure 8. Updated interferometric bench concept

A vertical architecture has been adopted to reduce the overall space required by the bench and reduce the risk of dust depositing on the detector. In this configuration, the detectors are oriented downwards, which is beneficial for sensors without protective windows, as their sensitive surface cannot be cleaned. The implementation of such a geometry introduces mechanical constraints related to gravity-induced flexure, vibration sensitivity, and thermal expansion. To ensure mechanical stability compatible with sub-pixel metrology requirements, low-expansion materials such as Invar are being considered for the main structural elements.

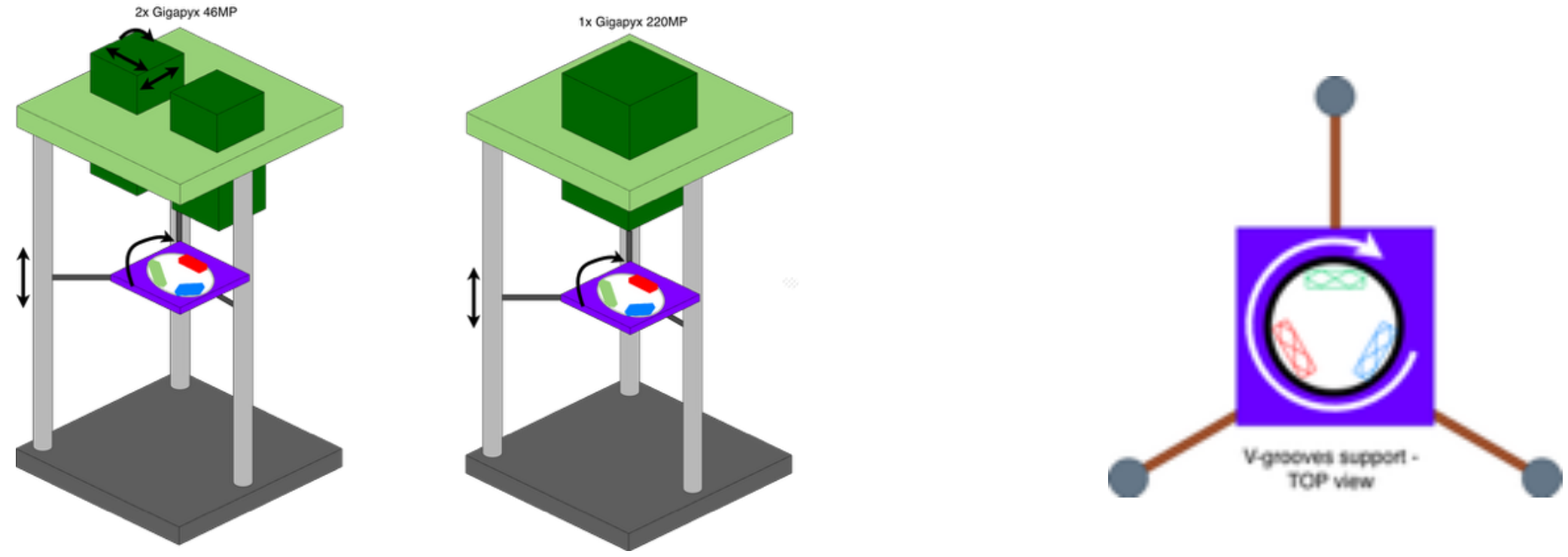

Figure 9. Left: view of the vertical bench with two 46-MP and one 220-MP detector. Right: top view of the V-groove support

In the dual-detector configuration, the two Pyxalis readout electronics are mounted side-by-side on a removable upper platform. One detector can be precisely moved relative to the other using motorized translation, tip-tilt, and rotation stages. These controlled adjustments introduce known detector misalignments, which are used to test calibration and centroiding algorithms and to determine acceptable alignment tolerances for the multi-detector focal planes. In the single-detector configuration, a single Pyxalis electronics unit is mounted on the upper platform to characterize the 220-MP sensor. The optical fiber assembly uses three v-groove mounts to generate interference fringes at multiple wavelengths, enabling the study of chromatic effects on calibration. The assembly can be manually translated vertically, along the optical axis, allowing the fiber to focal plane distance to be adjusted from 20 to 70 cm. The v-groove assembly can also be rotated to set and adjust the fringes orientation.

## 7. CONCLUSION AND PERSPECTIVES

This paper described the optical benches developed to characterize and calibrate CMOS detectors from the Pyxalis Gigapyx family for future high-precision astrometric missions such as HWO and Theia. A flat-field bench is used for electro-optical characterization, an interferometric bench for pixel geometry calibration, and the distortion bench based on the projection of simulated star field from an OLED source. Characterization results obtained with the 46-MP Gigapyx detector demonstrate promising performance for sub-micro-arcsecond astrometry. An upgraded interferometric bench is currently under development to support either two adjacent 46-MP detectors or a single 220-MP sensor, providing a representative platform for the calibration of future multi-detector focal-planes. Ongoing developments also target a space-compatible focal plane integrating four 220-MP detectors, dedicated space-qualified readout electronics, and a compact interferometric calibration module suitable for flight operation on future space mission such as HWO.

## ACKNOWLEDGEMENTS

The authors would like to thank all the researchers and engineers who contributed to the realization of this project but are not co-authors of this paper. This work was supported by the LabEx FOCUS ANR-11-LABX-0013, CNES, PYXALIS, and the French PEPR *Origine* programme (France 2030).

## REFERENCES

[1] Malbet, F., Boehm, C., Krone-Martins, A., et al. 2021, Experimental Astronomy, 51, 845
[2] Shao, M., Zhai, C., Nemati, B., et al. 2023, PASP, 135, 074502
[3] Ji, J.-H., Li, H.-T., Zhang, J.-B., et al. 2022, Research in Astronomy and Astrophysics, 22, 07200
[4] Nemati, B., Shao, M., Gonzalez, G., et al. 2020, in Society of Photo-Optical Instrumentation Engineers (SPIE) Conference Series, Vol. 11443, Space Telescopes and Instrumentation 2020: Optical, Infrared, and Millimeter Wave, ed. M. Lystrup & M. D. Perrin, 114430O
[5] Amiaux et al. 2026, in Society of Photo-Optical Instrumentation Engineers (SPIE): High-precision high-accuracy astrometry for the Habitable World Observatory
[6] Lizzana et al. 2026, in Society of Photo-Optical Instrumentation Engineers (SPIE): Calibrations for high precision differential astrometry onboard Theia and HWO
[7] Ardellier-Desages et al. 2026, in Society of Photo-Optical Instrumentation Engineers (SPIE): Technological maturation of a CMOS gigapixels astrometry instrument for Habitable World Observatory space mission
[8] Malbet, F., Labadie, L., Sozzetti, A., et al. 2022, in Society of Photo-Optical Instrumentation Engineers (SPIE) Conference Series, Vol. 12180, Space Telescopes and Instrumentation 2022: Optical, Infrared, and Millimeter Wave, ed. L. E.
[9] Rousset et al. 2026, in Society of Photo-Optical Instrumentation Engineers (SPIE): Calibration of pixel grid location of a CMOS silicon matrix detector based on Young's fringes for space-based high accuracy astrometry
[10] Crouzier, A. et al. A&A. 2016, 595, A108